\documentclass[pra,letterpaper,amsfonts]{revtex4}
\def\X{{x}}
\def\Y{{y}}
\def\tab{{Table }}
\def\ka{{k_1}}
\def\kb{{k_2}}
\def\kc{{k_3}}

\def\ke{{k_F}}
\def\kx{{k_i}}
\def\kg{{k_g}}
\usepackage{graphicx}

\usepackage{amsthm}
\usepackage{amsmath}
\usepackage{amsfonts}

\usepackage{amssymb}

\usepackage{bm}

\begin{document}

\bibliographystyle{apsrev}

\title{Interconversion of Nonlocal Correlations}
\author{Nick S. Jones}
\affiliation{Department of Mathematics, University of Bristol,
University Walk, Bristol, BS8 1TW, UK}
\author{Llu\'{\i}s Masanes}
\affiliation{Department of Mathematics, University of Bristol,
University Walk, Bristol, BS8 1TW, UK}

\begin{abstract}
In this paper we study the correlations that arise when two
separated parties perform  measurements on systems they hold
locally. We restrict ourselves to those correlations  with which
arbitrarily fast transmission of information is impossible. These
correlations are called nonsignaling. We allow the measurements to
be chosen from sets of an arbitrary size, but promise that each
measurement has only two possible outcomes. We find the structure
of this convex set of nonsignaling correlations by characterizing
its extreme points. Taking an information-theoretic view, we prove
that all of these extreme correlations are interconvertible. This
suggests that the simplest extremal nonlocal distribution (called
a PR box) might be the basic unit of nonlocality. We also show
that this unit of nonlocality is sufficient to simulate all
quantum states when measured with two outcome measurements.


\end{abstract}

\maketitle

\section{Introduction}
Measurements on parts of quantum states held by spatially
separated parties cannot be used for superluminal signalling; in
this respect quantum mechanics is a nonsignaling theory. John Bell
\cite{Bell} exposed a novel feature of the theory when he
considered a gedanken experiment of the following form: two
separated parties, Alice and Bob, locally measure two physical
systems which were, at an earlier time, very close together. Bell
found quantum states which display measurement outcome statistics
which vary, as Alice and Bob change their measurements, in a way
which cannot be explained by only assuming an exchange of
classical information when the two systems were close together in
the past. This behavior is termed quantum nonlocality and has
partial experimental validation (for a discussion of experimental
tests and loopholes see \cite{collinsthesis} and references
therein).

Quantum mechanics is not the only conceivable theory that predicts
correlations which, though they are nonsignaling, cannot be
understood as having been established in the past. This paper
investigates the structure of the set of all possible nonsignaling
correlations and attempts to characterize these in
information-theoretic terms.

Since quantum mechanics is so successful in its predictions, it
might seem unusual to consider other theories, with different
kinds of correlations, which are not physically instantiated.
There are practical and foundational physical motives, as well as
information-theoretic reasons, for considering a broader class of
correlations.

Motivated by the technological promise of quantum information,
there is a drive to understand the origins of quantum features
which may have concrete applications. They could be direct
consequences of the fact that quantum mechanics is a nonsignaling
theory or, alternatively, exploit other features of the theory.
Such concerns motivated the information-theoretic treatment of
nonlocal correlations in \cite{barrett}. A second reason for
interest in general nonsignaling theories is foundational; given
that quantum mechanics is a nonsignaling theory, what simplest
possible extra features must be added to explain the results of
our experiments? Popescu and Rohrlich \cite{popescu+rohrlich} show
that there exist nonsignaling correlations which cannot be
reproduced by quantum mechanics: why is quantum mechanics not the
most general kind of nonsignaling theory, what further constraints
does it satisfy?

In the context of communication complexity and cryptography,
interesting results have come from considering nonsignaling
correlations. Van Dam \cite{vandam2}
showed that, equipped with `superstrong nonlocal' correlations,
all bipartite communication complexity problems are rendered
trivial (requiring only one bit of communication). There has also
been work relating bit commitment to nonsignaling \cite{Wolf,Tony,
Buhrman}. In cryptography, it is best to have security proofs that
rely on a minimum number of principles; in \cite{qkd}, a key
distribution scheme is presented which can be proved secure by
only assuming nonsignaling.

Our work follows that of Barrett et al. \cite{barrett}. They
characterize the bipartite nonlocal correlations arising when
Alice and Bob can perform one of two measurements, each with an
arbitrary outcomes. They also provide results on the
interconversion of correlations and consider the case of more than
two parties.

\bigskip

In this paper, we consider the set of bipartite nonsignaling
correlations, where each party performs one from an arbitrary set
of measurements and each measurement has two possible outcomes (a
reversal of the situation in \cite{barrett}). This set is a convex
polytope and we characterize it in terms of its extreme points.
The structure of these extreme points has already been used by one
of the authors \cite{acin} to show that, for all nonsignaling
theories, the more incompatible two observables are, the more
uncertain their corresponding outcomes. From an
information-theoretic perspective, we also prove that all nonlocal
extremal correlations are interconvertible, in the sense that
given sufficient copies, any one can simulate any other.
Consequently, any nonlocal extremal distribution can simulate any
non-extremal one. The simplest extremal nonlocal correlations are
called Popescu-Rohrlich (PR) boxes \cite{barrett}. One can thus
consider the PR box as the unit resource of bipartite nonlocal
correlations, in the same fashion as the singlet is considered the
unit resource of quantum correlations. It is, as yet, unclear if
they can serve as s sufficient unit in more general cases. Since
quantum correlations are nonsignaling, all those within the
polytope considered  can be simulated by PR boxes. It has
previously been shown \cite{Cerf} that all possible projective
measurements on the singlet state of two qubits can be simulated
using just one PR box and shared randomness (our result can be
seen as an extension from projective measurements on singlets to
POVMs on general bipartite quantum states).

\bigskip

This paper is structured in the following way. In Section II the
set of nonsignaling correlations is characterized in terms of
inequalities. Section III reviews past results and characterizes
the structure of this set in terms of its extreme points. Section
IV is devoted to the inter-convertibility of nonsignaling
correlations. Section V concludes and shows, by giving an example,
that the extreme points in more general cases have nonuniform
marginals and thus lack the simple structures found in Section III
and in \cite{barrett}.

\section{No-signalling correlations}

In what follows, we will consider two parties ---Alice and Bob---
each performing space-like separated operations. Each possesses a
physical system which can be measured in several distinct ways and
each measurement can yield several distinct results. Let $\X$
($\Y$) denote the observable chosen by Alice (Bob) (these will
also be called {\em inputs}), and $a$ ($b$) be the result of
Alice's (Bob's) measurement (these will also be called {\em
outputs}). The statistics of these measurements define a joint
probability distribution for the outputs, conditioned on the
inputs, $P_{ab|\X\Y}$, which satisfies the usual constraints:
\begin{eqnarray}
&& P_{ab|\X\Y}\geq 0\;\; \forall\; a,b,\X,\Y, \label{1}\\
&& \sum_{a,b}P_{ab|\X\Y}=1\;\; \forall \; \X,\Y. \label{2}
\end{eqnarray}
We consider the input $\X$ ($\Y$) to take values from an alphabet
of length $d_\X$ ($d_\Y$), that is, $\X\in \{0,...,d_\X-1\}$ and
$\Y\in \{0,...,d_\Y-1\}$. The output $a$ ($b$) takes values from
an alphabet of length $d_a$ ($d_b$), $a\in \{0,...,d_a-1\}$ and
$b\in \{0,...,d_b-1\}$.

\subsection{No-signalling constraints}
The requirement that Alice and  Bob  cannot signal to each other
by using their correlations is equivalent to the condition that
Alice's output is independent of
 Bob's input, $P_{a|\X}$ is independent of $\Y$ (and vice-versa):
\begin{eqnarray}
&& \sum_{b}P_{ab|\X\Y}=\sum_{b}P_{ab|\X\Y'}\;\; \forall \; a,\X,\Y,\Y',\label{3}\\
&& \sum_{a}P_{ab|\X\Y}=\sum_{a}P_{ab|\X'\Y}\;\; \forall \;
b,\X,\X',\Y.\label{4}
\end{eqnarray}
For fixed $d_\X$, $d_\Y$, $d_a$, and $d_b$, the set of probability
distributions Eqs. (\ref{1}, \ref{2}) is convex and has a finite
number of extreme points. In other words, it is a convex polytope.
It is known that the intersection of a polytope with an affine
set, like the one defined by the no-signaling constraints
(\ref{3}, \ref{4}), defines another convex polytope. From now on,
all distributions are assumed to belong to this set. In this paper
such distributions are represented by  tables of the form given in
\tab I.

\begin{table}
\begin{center}
\begin{tabular}{|ll||ll|ll|ll|ll|} \hline
   &$x$&$0$ & &$1$& & $\hdots$& & $d_\X-1$& \\
   $y$&&&&&&&&&\\\hline\hline
   $0$ &&$P_{00|00}$&$ P_{10|00}$  & $P_{00|10}$&$ P_{10|10}$  &$ $& &$P_{00|d_\X-1,0}$&$ P_{10|d_\X-1,0}$   \\
   &&$P_{01|00}$&$ P_{11|00}$  & $P_{01|10}$&$ P_{11|10}$  &$ $& &$P_{01|d_\X-1,0}$&$ P_{11|d_\X-1,0}$  \\\hline
   $1$&&$P_{00|01}$&$ P_{10|01}$  & $P_{00|11}$&$ P_{10|11}$  &$ $& &$P_{00|d_\X-1,1}$&$ P_{10|d_\X-1,1}$  \\
  &&$P_{01|01}$&$ P_{11|01}$  & $P_{01|11}$&$ P_{11|11}$  &$ $& &$P_{01|d_\X-1,1}$&$ P_{11|d_\X-1,1}$ \\\hline
   \vdots&&&&&&$\ddots$&&&\\\hline
$d_\Y-1$&&$P_{00|0,d_\Y-1}$&$ P_{10|0,d_\Y-1}$  & $P_{00|1,d_\Y-1}$&$ P_{10|1,d_\Y-1}$  &$ $& &$P_{00|d_\X-1,d_\Y-1}$&$ P_{10|d_\X-1,d_\Y-1}$  \\
  &&$P_{01|0,d_\Y-1}$&$ P_{11|0,d_\Y-1}$  & $P_{01|1,d_\Y-1}$&$ P_{11|1,d_\Y-1}$  &$ $& &$P_{01|d_\X-1,d_\Y-1}$&$ P_{11|d_\X-1,d_\Y-1}$ \\\hline

\end{tabular}
\end{center}
 \caption{ \label{re} This table represents a general probability distribution for $d_\X$ and $d_\Y$ arbitrary and $d_a=d_b=2$.
 The distribution is broken into $d_\X\times d_\Y$ cells, with one cell for every input pair $(\X,\Y)$.
 Each cell specifies the probabilities of the four possible outcomes given these inputs (these must sum to one).
 The nonsignaling conditions require, for example, that $P_{00|00}+P_{01|00}=P_{00|01}+P_{01|01}$ and that
 $P_{01|00} + P_{11|00}=P_{01|d_\X-1,0} + P_{11|d_\X-1,0}$.}
\end{table}

\subsection{Local correlations}
Local correlations are those that can be reproduced by parties
equipped only with shared randomness. These are a proper subset of
 nonsignaling correlations. One can always write them as:
\begin{equation}
P_{ab|\X\Y}=\sum_e p_e P_{a|\X e}P_{b|\Y e}. \label{5}
\end{equation}
A protocol for generating $P_{ab|\X\Y}$ is the following: With
probability $p_e$ Alice (Bob) samples from the distribution
$P_{a|Xe}$ ($P_{b|Ye}$).  It is known that the set of local
correlations is a convex polytope with some of the facets being
Bell-like inequalities \cite{Peres}. The extreme points of this
polytope correspond to local deterministic distributions, that is
$P_{ab|\X\Y}=\delta_{a,f(\X)} \delta_{b,g(\Y)}$, where $f(\X)$ and
$g(\Y)$ map each input value to a single output value.
Correlations that are not of the form (\ref{5}) are called
nonlocal.

\subsection{Quantum Correlations}

Quantum correlations are generated if Alice and  Bob share quantum
entanglement. These can be written as:
\begin{equation}
P_{ab|\X\Y}=\mbox{tr}[F_a^\X\otimes F_b^\Y \rho], \label{6}
\end{equation}
where $\{F_0^\X,...,F^\X_{d_a-1}\}$,
$\{F_0^\Y,...,F^\Y_{d_b-1}\}$, are positive operator valued
measures for each $\X$ and $\Y$, and $\rho$ is a density matrix.
Though this set is convex, it is not a polytope; it includes all
local correlations and also probability distributions which are
nonlocal. It is, however, smaller than the full set of
nonsignaling correlations. This was proved in
\cite{popescu+rohrlich} by providing an example of a nonsignaling
distribution forbidden by quantum mechanics.

\section{Extreme nonsignaling correlations}
The full set of extremal distributions for the general situation
where $d_\X$, $d_\Y$, $d_a$ and $d_b$ is not yet characterized. In
what follows, previous work considering the case where
$d_a=d_b=d_\X=d_\Y=2$ and the case for $d_\X=d_\Y=2$, and both
$d_a$ and $d_b$ arbitrary will be reviewed \cite{barrett,cir}.
Next, the extreme points for $d_a=d_b=2$ and both $d_\X$ and
$d_\Y$ arbitrary will be presented.

\subsection{Reversible local transformations}
Applying reversible local transformations to a distribution does
not change its nonlocal properties. We say that two distributions
are {\em equivalent} if one can be transformed into the other by
means of local reversible transformations. Identifying {\em
classes} of extremal distributions which are equivalent simplifies
the task of categorizing all of them. It is sufficient to quote
one representative element from each equivalence class. Let us
list all possible local reversible transformations:
\begin{itemize}
    \item Permute the ordered set of input values for each party,
    $(0,1,\ldots d_\X-1)$ and $(0,1,\ldots d_\Y-1)$.

    \item Permute the ordered set of output values depending on the input.
    To indicate that $a$ and $b$ are associated with the particular inputs
    $(\X,\Y)$, the notation $a_\X$ and $b_\Y$ will sometimes be
    used. Summarizing, one can apply a different permutation to
    each of the $a_\X$  $(b_\Y)$, for each value of $\X$ $(\Y)$.
\end{itemize}

\subsection{Binary inputs and outputs}
All extremal nonlocal distributions for the case
$d_\X=d_\Y=d_a=d_b=2$ are equivalent to:

\begin{equation}\label{8.5}
\begin{tabular}{|ll||ll|ll|} \hline
   &$x$&$0$\;\;\;\;\;\;\;& &$1$\;\;\;\;\;\;\;&  \\
   y&&&&&\\
  \hline\hline $0$&&$1/2$ &$0$ &$1/2$ &$0$\\
  &&$0$ &$1/2$ &$0$ &$1/2$\\\hline

 $1$&&$1/2$ &$0$ &$0$ &$1/2$\\
  &&$0$ &$1/2$ &$1/2$ &$0$\\\hline

\hline
\end{tabular}
\end{equation}

(this format is explained in \tab I) or alternatively:

\begin{equation}\label{9}
p_{ab|\X\Y} = \left\{ \begin{array}{r@{\quad: \quad}l} 1/2 &
a+ b\mod 2=\X \Y\\
0 & \mbox{otherwise,}
\end{array} \right.
\end{equation}
where it is understood that $a$ and $b$ are locally uniformly
distributed. This distribution is also called a PR box and
constitutes the paradigm of nonlocality. PR boxes have their
outputs together, depending on their inputs together; but they are
nonsignaling since their outputs are (locally) random, obeying
 Eqs. (\ref{3},\ref{4}).

\subsection{Binary inputs and arbitrary outputs}
Barrett et al. \cite{barrett} provided the following
characterization for the case where $d_\X=d_\Y=2$ and arbitrary
$d_a,\;d_b$ outcomes. Each inequivalent extremal nonlocal
distribution is characterized by one value of the parameter
$k\in\{2,\ldots \min (d_{a},d_{b})\}$. For each $k$, its
corresponding distribution is
\begin{equation}\label{10}
p_{ab|\X\Y} = \left\{ \begin{array}{r@{\quad}c@{\quad}l} 1/k &:&
(b-a) \mathrm{\ mod\ } k = \X \Y\\
\\
0 &:& \mathrm{otherwise,}
\end{array} \right.
\end{equation}
where $a,b\in\{0,\ldots,k-1\}$ and are locally uniformly
distributed. Note that Eq. (\ref{9}) is recovered when
$d_a=d_b=2$.

\subsection{Arbitrary inputs and binary outputs}
In what follows, one of the main results of our paper is
presented. We give a characterization of all extreme distributions
for the case where $d_\X$ and $d_\Y$ are arbitrary, and
$d_a=d_b=2$. The proof of this result is provided in the appendix.

\bigskip
{\bf Result 1:} {\em \tab II provides at least one representative
element of all classes of extremal correlations for a given $d_\X$
and $d_\Y$. Each of these distributions is characterized as
follows:}
    \begin{enumerate}
    \item {\em Giving  two integers $g_\X$ and $g_\Y$, where $g_\X \in\{2,3,\cdots
d_\X\}$ and $g_\Y \in\{2,3,\cdots d_\Y\}$ if the distribution is
nonlocal, and $g_\X=g_\Y=0$ if the distribution is local.}
    \item {\em And assigning perfect correlation or
    anti-correlation to all the cells with a question mark `$\,?$',
    that is}
\begin{equation}\label{pio}
    \begin{tabular}{|cc|}\hline
      ? & \hspace{.8cm} \\
        &  \\
    \hline\end{tabular}
    \quad=\quad
\begin{tabular}{|cc|}\hline
      1/2 & 0 \\
      0 & 1/2 \\
    \hline\end{tabular}
    \quad \mbox{or} \quad
    \begin{tabular}{|cc|}\hline
      0 & 1/2 \\
      1/2 & 0 \\
    \hline\end{tabular}\ .
\end{equation}
    \end{enumerate}
\bigskip

\begin{table}[h!]
\begin{center}
\begin{tabular}{|ll||ll|ll|ll|ll|ll|ll|ll|ll|} \hline
   &$x$&$0$\;\;\;\;\;\;\;& &$1$\;\;\;\;\;\;\;& & $2$\;\;\;\;\;\;\;& &$\hdots$ & &$g_\X-1$& &$g_\X$\;\;\;\;\;\;\;& &$\hdots$  & & $d_\X-1$ &  \\
   y&&&&&&&&&&&&&&&&&\\
  \hline\hline $0$&&$1/2$ &$0$ &$1/2$ &$0$&$1/2$ &$0$ & $ $ & &$1/2$ &$0$&$1/2$ &$0$& $ $& &$1/2$ &$0$\\
  &&$0$ &$1/2$ &$0$ &$1/2$&$0$ &$1/2$ & $ $ & &$0$ &$1/2$&$1/2$ &$0$& $ $ & &$1/2$
 &$0$\\\hline

 $1$&&$1/2$ &$0$ &$0$ &$1/2$&$?$ &$ $ & & & ? &$ $&$1/2$ &$0$& & &$1/2$ &$0$\\
  &&$0$ &$1/2$ &$1/2$ &$0$&$ $ & & & &$ $ &&$1/2$ &$0$& &
  &$1/2$&0\\\hline

  $2$&&$1/2$ &$0$ &? &$ $&? &$ $ &$ $ & &? &$ $&$1/2$ &$0$&$ $ & &$1/2$ &$0$\\
  &&$0$ &$1/2$ &$ $ & &  &  &$ $ & &  & &$1/2$ &$0$&$ $ &
  &$1/2$&0\\\hline
$\vdots$& &&& & & &&$\ddots$&&&&&&$\ddots$&&&\\\hline

  $g_\Y-1$&&$1/2$ &$0$ &? &$ $&? &$ $ &$ $ & &? &$ $&$1/2$ &$0$&$ $ & &$1/2$ &$0$\\
  &&$0$ &$1/2$ &$ $ & &  &  &$ $ & &  & &$1/2$ &$0$&$ $ &
  &$1/2$&0\\\hline

  $g_\Y$&&$1/2$ &$1/2$ &$1/2$ &$1/2$&$1/2$ &$1/2$ & & &$1/2$ &$1/2$&$1$ &$0$& & &$1$ &$0$\\
  &&$0$ &$0$ &$0$ &$0$&$0$ &$0$ & & &$0$ &$0$&$0$ &$0$& &
  &$0$&0\\\hline

$\vdots$&& && & & &&$\ddots$&&&&&&$\ddots$&&&\\\hline

  $d_\Y-1$&&$1/2$ &$1/2$ &$1/2$ &$1/2$&$1/2$ &$1/2$ & & &$1/2$ &$1/2$&$1$ &$0$& & &$1$ &$0$\\
  &&$0$ &$0$ &$0$ &$0$&$0$ &$0$ & & &$0$ &$0$&$0$ &$0$& &
  &$0$&0\\\hline

\hline
\end{tabular}
\end{center}
 \caption{\label{r} This table gives a representative element of all
classes of extreme points, where Alice (Bob) has $d_\X$ ($d_\Y$)
different input settings, and $g_\X$ ($g_\Y$) of them are
nondeterministic. Cells containing a `?' can either be perfectly
correlated (like the cell corresponding to $\X=\Y=0$) or
anti-correlated (like the cell corresponding to $\X=\Y=1$).}
\end{table}

As one can see in \tab II, each party has two kinds of input
settings: (i) the deterministic ones ($\X\geq g_\X$ for Alice)
have a fixed outcome, (ii) the nondeterministic ones ($\X<g_\X$
for Alice) have uniform probabilities for their corresponding
outcomes, $P_{0|\X}=P_{1|\X}=1/2$. There are $g_\X$
nondeterministic input settings and $d_\X-g_\X$ deterministic
input settings in Alice's site and analogously for Bob.  The
representative distributions are chosen to have the outcomes for
all deterministic input settings fixed to `$0$'.

The following observation will prove crucial. When the
distribution is nonlocal, that is $g_\X,g_\Y\geq 2$, there is
always a PR box structure when both parties restrict to $x,y\in
\{0,1\}$.

Extreme points for which $g_\X=d_\X$ and $g_\Y=d_\Y$ can be
algebraically characterized  by: $a_x +  b_y =
\delta_{\X,1}\delta_{\Y,1}+\sum_{(i,j)\in Q}
\delta_{\X,i}\delta_{\Y,j}\;\; \mbox{mod 2}$. Here  $Q$ is  any
subset of the set $\{1,...,d_x\}\times \{1,...,d_y\} - \{(1,1)\}$.

\section{Interconversion of nonlocal correlations}

In this section we prove that all extremal nonlocal correlations
with binary outputs can be interconverted. This means that all
contain the same kind of nonlocality. By saying that the
distribution $P_{ab|\X\Y}$ can be converted into $P'_{ab|\X\Y}$ we
mean: given enough copies (realizations) of $P_{ab|\X\Y}$, Alice
and Bob can simulate the statistics of $P'_{ab|\X\Y}$ for any
value of $\X$ and $\Y$ that they independently choose. We assume
that the two parties can perform local operations and have
unlimited shared randomness. This is a fair assumption because
with these resources (shared randomness and local operations) we
cannot create nonlocality.

\bigskip
{\bf Result 2:} {\em All  nonlocal extremal correlations with
arbitrary $d_\X$ and $d_\Y$, and binary output ($d_a=d_b=2$) are
interconvertible.}
\bigskip

In order to prove this statement, we first argue that all extremal
nonlocal correlations can simulate a PR box, and second, we prove
that PR boxes are sufficient to simulate all extremal
distributions. By recalling that all distributions can be written
as  probabilistic mixtures of extreme points (noting that such
mixtures can be reproduced by shared randomness), one can also
make the following statement:

\bigskip
{\bf Result 3:} {\em PR boxes are sufficient to simulate all
nonsignaling correlations with binary output ($d_a=d_b=2$).}
\bigskip

By looking at \tab II one can see that, if Alice and Bob share a
nonlocal distribution ($g_\X,g_\Y\geq 2$), they have a PR box when
restricting $\X,\Y\in\{0,1\}$. This shows that a single copy of
any nonlocal extremal distribution can simulate a PR box. Next, we
present a protocol that allows Alice and Bob to simulate any
distribution of the form described in \tab II, by only using a
finite number of PR boxes (\ref{9}). This protocol is based on an
idea presented in \cite{vandam2}.

If $g_\X=g_\Y=0$ the distribution is local, and thus, it can be
simulated with the protocol detailed in Section II.B without using
PR boxes. When $g_\X, g_\Y\geq 2$, however, such protocols cannot
be used. Let us first describe how to make the simulation when
Alice and Bob choose inputs $\X\leq g_\X-1$ and $\Y\leq g_\Y-1$.
Within this range of input settings the outcomes $a_\X$ and $b_\Y$
are locally random and they are either perfectly correlated
($a_\X+ b_\Y=0 \mod 2$), or anti-correlated ($a_\X+ b_\Y=1 \mod
2$). Equivalently,  any distribution of the form defined by \tab
II for inputs $\X\leq g_\X-1$ and $\Y\leq g_\Y-1$ is equally well
defined by a function:

\begin{equation}\label{myfriendgoo}
F(\X,\Y)=a_\X + b_\Y .
\end{equation}
Throughout this section all equalities are always modulo 2 and
thus we omit the specification `(mod 2)'. Let us expand $\X$ and
$\Y$ in binary: $\X=(\X_1\X_2\ldots \X_{n_\X})$,
$\Y=(\Y_1\Y_2\ldots \Y_{n_\Y})$, where $n_\X=\lceil \log_2
g_\X\rceil$ and $n_\Y= \lceil \log_2 g_\Y\rceil$. The function
$F(\X,\Y)$ can always be expressed as a polynomial of the binary
variables $\X_1,\ldots \X_{n_\X},\Y_1,\ldots \Y_{n_\Y}$. More
specifically, one can always write $F(\X,\Y)$ as a finite sum of
products
\begin{equation}\label{}
    F(\X,\Y)=\sum_{i=1}^{2^{n_\Y}}P_i(\X) Q_i(\Y),
\end{equation}
where each $P_i(\X)$ is a polynomial in the variables
$\{\X_1,\X_2,...,\X_{nx}\}$, and each $Q_i$ is a monomial in the
variables $\{\Y_1,\Y_2,...,\Y_{n_\Y}\}$. The sum has at most
$2^{n_y}$ terms, because there are $2^{n_\Y}$ distinct monomials
in the variables $\{\Y_1,\Y_2,...,\Y_{n_\Y}\}$.

\bigskip

Let us describe the Protocol. Suppose Alice and Bob choose the
input settings $\X\leq g_\X-1$ and $\Y\leq g_\Y-1$. Alice (Bob)
evaluates the $2^{n_y}$ numbers $r_i=P_i(\X)$ ($s_i=Q_i(\Y)$)
depending on the $\X$ ($\Y$) chosen. Then, Alice (Bob) inputs the
binary number $r_i$ ($s_i$) in the $i^{\mbox{\scriptsize th}}$ PR
box and obtains the outcome $a_i$ ($b_i$). They do such operations
for all $i=1,\ldots 2^{n_\Y}$. Finally, each party computes its
output of the simulated distribution $(a_\X,b_\Y)$ by summing the
local outputs of the PR boxes:
\begin{equation}\label{}
    a_\X:=\sum_{i=1}^{2^{n_\Y}} a_i \quad\quad\quad\quad
    b_\Y:=\sum_{i=1}^{2^{n_\Y}} b_i.
\end{equation}
The protocol works because of the next chain of equalities:
\begin{equation}
F(\X,\Y)=\sum_{i=1}^{2^{n_\Y}}P_i(\X)
Q_i(\Y)=\sum_{i=1}^{2^{n_\Y}}r_is_i=\sum_{i=1}^{2^{ny}} \left(a_i
+ b_i\right)=\sum_{i=1}^{2^{ny}}a_i + \sum_{j=1}^{2^{ny}}b_j=a_\X+
b_\Y.
\end{equation}
To see the third equality, just recall that for each PR box $a_i+
b_i=r_i s_i$ holds.

\bigskip

 Let us now consider the case where Alice picks an input
$\X\geq g_\X$, she must then assign to $a_\X$ the corresponding
deterministic value and analogously for Bob. One can see that the
simulation protocol works for all values of $\X$ and $\Y$.

\bigskip

A corollary of Result 3 is the following. Since quantum
correlations are nonsignaling, the statistics of any two-outcome
measurements experiment on any bipartite quantum state, can also
be simulated with PR-boxes (as noted in the introduction, this
result extends \cite{Cerf}).

\section{Discussion}

In this paper we have given a complete characterization of the
extremal nonsignaling bipartite probability distributions with
binary outputs. We have grouped them into equivalence classes
under local reversible transformations. One can see in \tab II
that these extremal distributions have a more complicated
structure than in the  binary input scenario (\ref{9},\ref{10}).
Nevertheless, if we consider purely nonlocal distributions
($g_x=d_x$ and $g_y=d_y$) all the marginals are unbiased
($P_{a|x}=P_{b|y}=1/2$) and  they are easily defined by specifying
which input pairs $(x,y)$ have correlated outputs and which
$(s,y)$ have anti-correlated outputs. In  more general cases the
extremal distributions stop showing these simple symmetries. An
example of this more complex structure is given in
 \tab III. This is  an extremal distribution for the case
$d_a=d_b=d_x=d_y=3$, which we discovered numerically (that this is
extremal can be verified by using arguments similar to those in
Part 3 of the Appendix). Its corresponding marginals are not
unbiased and there are also some input pairs, $(x,y)$, for which,
once the output of one party is fixed the outcomes of the other
remain uncertain.

\begin{table}
\begin{center}
\begin{tabular}{|ll||lll|lll|lll|} \hline
   &\; $x$&$0$&&&$1$&&&2&&\\
$y$ \;&&&&&&&&&&\\ \hline\hline
      $0$&&1/4&0&1/4&1/2&0&0&1/2&0&0 \\
      &&0&1/4&0&0&1/4&0&0&1/4&0\\
       &&1/4&0&0&0&0&1/4&0&0&1/4\\\hline
$1$&&1/2&0&0&1/4&1/4&0&1/4&0&1/4 \\
&&0&1/4&0&1/4&0&0&1/4&0&0\\
&&0&0&1/4&0&0&1/4&0&1/4&0\\\hline
$2$&&1/2&0&0&1/4&1/4&0&0&1/4&1/4 \\
&&0&1/4&0&1/4&0&0&1/4&0&0\\
&&0&0&1/4&0&0&1/4&1/4&0&0\\
\hline
\end{tabular}

\end{center}
\caption{\label{rel} An extreme point of the nonsignaling polytope
for 3 input settings each with 3 possible outcomes. Each cell
contains 9 probabilities associated with the $3\times 3$ possible
outcome pairs.}
\end{table}

We have shown that all extremal nonlocal distributions with binary
outputs are interconvertible. We have also given a specific
protocol to implement this interconversion. By looking at the
asymmetric structure of the extremal distribution in  \tab III,
one sees that this protocol is not directly applicable in general.
In particular it is an open question whether this distribution can
be simulated by PR boxes. We conclude by noting that, just as
treating the singlet as a unit of entanglement motivated numerous
resource based questions (asymptotic interconversions,
multipartite scenarios, etc) so too there is an analogous set of
unanswered information-theoretic problems involving units of
nonlocality.

{\em Note added.} After the completion of this work, the authors
were made aware that similar results have been obtained by J.
Barrett and S. Pironio \cite{Pironio}.

\section*{Acknowledgments}

The authors would like to thank J. Barrett, S. Popescu and T.
Short for discussions. This work has been supported by the U.K.
Engineering and Physical Sciences Research Council (IRC QIP).

\section*{APPENDIX}
In this appendix we give the proof of Result 1. Firstly we show
that any nonsignaling distribution can be expressed as a convex
combination of distributions equivalent to  ones of the form given
in \tab II. Secondly we show that all distributions of the form
given in \tab II are extremal.

\bigskip

Some simple definitions will prove useful throughout this
appendix. The word `cell' refers to the set of four outcome
probabilities $P_{00|\X\Y}$, $P_{10|\X\Y}$, $P_{01|\X\Y}$,
$P_{11|\X\Y}$ associated with the input pair $(\X,\Y)$. It will be
useful to think of $P_{ab|\X\Y}$ as a table of cells with $d_\X$
columns and $d_\Y$ rows where, associated with each entry of the
table, $(\X,\Y)$, there is a cell of four probabilities (See \tab
\ref{re}). We say that a cell  has `one zero' if it has at least
one of the four entries it contains set to zero. 
We call $P_{a|\X}, P_{b|\Y}$ $\forall a,b,\X,\Y$ the `marginals'.
Specifically we define $P_{a=0|\X=i}\equiv l_i$ and
$P_{b=0|\Y=i}\equiv m_i$. We now sketch the  strategy adopted for
Parts 1 and 2 of the proof.

\bigskip

An arbitrary distribution $P^{(1)}$ is expressed  as a convex
combination of two distributions:
\begin{equation}\label{}
    P^{(1)}= \lambda_1 P^{(1)}_1 + (1-\lambda_1) P^{(1)}_2.
\end{equation}
We require that the new distributions, $P^{(1)}_1$ and
$P^{(1)}_2$, have one more entry of their tables set equal to
zero. Next we select one of them: $P^{(1)}_1$ or $P^{(1)}_2$. We
then repeat
 the above decomposition for the selected distribution. A schematic of the approach
 is:
\begin{eqnarray}
P^{(1)}&=& \lambda_1 P^{(1)}_1 + (1-\lambda_1) P^{(1)}_2, \label{A1}\\
P^{(2)}&=&  P^{(1)}_{\ka}, \quad \ka\in\{1,2\},\label{or1}\\
P^{(2)}&=& \lambda_2 P^{(2)}_1 + (1-\lambda_2) P^{(2)}_2,\label{A3}\\
P^{(3)}&=&  P^{(2)}_{\kb}, \quad \kb\in\{1,2\},\label{or2} \\
P^{(3)}&=& \lambda_3 P^{(3)}_1 + (1-\lambda_3) P^{(3)}_2,\label{A4.5}\\
\vdots \nonumber\\
P^{(F)}&=& \lambda_F P^{(F)}_1 + (1-\lambda_F) P^{(F)}_2,
\label{A5}
\end{eqnarray}
where the $P^{(i)}$ are probability distributions $P_{ab|\X\Y}$
expressed as vectors and $\lambda_i\in [0,1]$. At each step a
distribution with one more entry set to zero is selected. It may
happen that a distribution $P^{(i)}$ will already have a zero at
the position demanded in the next step (e.g. if $P^{(i)}$ is
already extremal). The expression $k\in\{1,2\}$ (e.g.  in Eqs.
(\ref{or1},\ref{or2})) indicates that the consecutive steps of the
proof hold independently of which of the two distribution is
chosen. The procedure stops when the new distribution chosen,
$P^{(F)}_{k_F}$, is equivalent to one of the form given in \tab
II. We will see that this procedure, based on successive zeroing
of entries, always finishes.

\bigskip

The proof of Result 1 is in three parts.
\begin{itemize}
    \item In Part $1$ we show that any probability distribution can be
expressed as a convex combination of probability distributions
which have at least one zero in every cell and which have the same
marginals ($P_{a|\X}$, $P_{b|\Y}$, $\forall\, a,b,\X,\Y$) as the
original distribution.
    \item In Part $2$ we show that any probability distribution with one
zero in every cell can be expressed as convex combinations of
probability distributions which are locally equivalent to \tab
\ref{r}.
    \item In Part $3$ we show that all distributions of the form defined in
\tab \ref{r} are extremal.
\end{itemize}

\subsection*{Part 1}
This Part is broken in two. We first show  that every cell  can be
expressed as a convex combination of two cells which satisfy the
following constraints. Each has the same marginals as the original
cell and also has at least one of their four entries set to zero.
The nonsignaling conditions (\ref{3},\ref{4}) mean that all cells
in the same column, $\X$, have the same marginals
$P_{a=0|\X}=l_x$, $P_{a=1|\X}=1-l_x$. Consider a cell with
marginals $l_\X\geq m_\Y \geq \frac{1}{2}$. Given $l_\X$ and
$m_\Y$, there is one free parameter, $c$, needed to completely
specify the cell $(x,y)$:
\begin{equation}\label{pipi}
\begin{tabular}{|ccc|}\hline
      $c$ && $m_\Y-c$ \\
       &\;\;\;\;& \\
      $l_x-c$ && $1+c-m_y-l_x$ \\
    \hline\end{tabular}
    _{\;\X\Y.}
\end{equation}
By the above notation we mean:
$P_{00|\X\Y}=c;\;P_{10|\X\Y}=m_\Y-c;\;P_{01|\X\Y}=l_\X-c;\;P_{11|\X\Y}=1+c-m_\Y-l_\X$.

If we require the positivity of the four elements of the cell,
then $c\in [m_\Y+l_\X -1, m_\Y]$. One can readily check that all
cells with allowed values of $c$ can be written as convex
combinations of the two cells where $c=m_\Y$ and $c=m_\Y +l_\X
-1$:

\begin{equation}\label{pio}
    \begin{tabular}{|ccc|}\hline
      $c$ && $m_y-c$ \\
       &\;\;& \\
      $l_x-c$ && $1+c-m_y-l_x$ \\
    \hline\end{tabular}
    _{\;\X\Y}
  =\lambda\,
\begin{tabular}{|ccc|}\hline
      $m_y$ && $0$ \\
       &\;\;\;\;\;\;\;\;& \\
      $l_x-m_y$ && $1-l_x$ \\
    \hline\end{tabular}
    _{\;\X\Y}
  +(1-\lambda)\,
    \begin{tabular}{|ccc|}\hline
      $m_\Y +l_\X -1$ && $1-l_x$ \\
       &\;\;& \\
      $1-m_y$ && $0$ \\
    \hline\end{tabular}
    _{\;\X\Y.}
\end{equation}

Instead of $l_\X\geq  m_\Y \geq \frac{1}{2}$, cells can satisfy
different inequalities e.g. $m_\Y\geq  l_\X \geq \frac{1}{2}$ or
$m_\Y\geq \frac{1}{2} \geq  l_\X$. Using  symmetries, one can see
that, whatever inequalities are satisfied, any cell can be
expressed as a convex combination of two, one zero, cells in a
similar manner.

\bigskip

In the second half of this Part we generalise from single cells to
the whole distribution. An iterative procedure of the form
described in Eqs. (\ref{A1}-\ref{A5}) can  be applied. Any
distribution, $P^{(1)}$, can be expressed as a convex combination
of two distributions, $P^{(1)}_1$ and $P^{(1)}_2$. Both
$P^{(1)}_1$ and $P^{(1)}_2$ have all cells equal to $P^{(1)}$,
except for the cell $(\X=0,\Y=0)$. This cell has the same
marginals as in $P^{(1)}$ but also has one more zero. $P^{(1)}_1$
(or $P^{(1)}_2$) can again be expressed as a convex combination of
two distributions $P^{(2)}_1$ and $P^{(2)}_2$ which are identical
to $P^{(1)}_1$ (or $P^{(1)}_2$) - they also have $(\X=0,\Y=0)$ as
a one zero cell - except that they also both have $(\X=0,\Y=1)$ as
a one zero cell (with the same marginals). This procedure can be
extended to all cells until the final step has a probability
distribution which is a convex combination of two distributions
which have at least one zero in every cell. It follows that any
probability distribution can be expressed as a convex combination
of probability distributions which have at least one zero in every
cell and which have the same marginals  as the original
distribution.

\subsection*{Part 2}
In this part we show that distributions with one zero in every
cell can be expressed as a convex combination of distributions
equivalent to \tab II. The argument exploits the fact that the
only parameters describing distributions with one zero in every
cell are their marginals. It considers first a cell ($1$), then a
column ($2$) and finally a generic table ($3$).

\subsubsection{A Cell}
In this section we identify the constraints on the marginals in
one zero cells. Consider a cell in the first column $(\X=0, \Y=i)$
with the form:

\begin{equation}\label{A6}
\begin{tabular}{|ccc|}\hline
      $m_i$ && $0$ \\
       &\;\;\;\;& \\
      $l_0-m_i$ && $1-l_0$ \\
    \hline\end{tabular}
    _{\;0i.}
\end{equation}
Note that the marginals are $P_{a=0|\X=0}=l_0$,
$P_{b=0|\Y=i}=m_i$.
Positivity requires that $l_0\in [m_i,1]$. 
For the same marginals, $l_0$ and $m_i$, if $P_{00|0i}=0$, instead
of $P_{10|0i}=0$:
\begin{equation}\label{A6}
\begin{tabular}{|ccc|}\hline
      $0$ && $m_i$ \\
       &\;\;\;\;& \\
      $l_0$ && $1-l_0-m_i$ \\
    \hline\end{tabular}
    _{\;0i,}
\end{equation}
then $l_0\in [0,1-m_i]$. If $P_{01|0i}=0$ instead then $l_0\in
[0,m_i]$. Finally, if $P_{11|0i}=0$ then $l_0\in [1-m_i,1]$. In an
arbitrary one zero cell, $l_0$ will thus lie in one of four
ranges:
\begin{equation}
[0,1-m_i],\quad \quad [0,m_i],\quad \quad [m_i,1],\quad \quad
[1-m_i,1], \label{A8}\end{equation} depending on which of its four
elements is zero. Part of the information in Eq. (\ref{A8}) can be
 expressed as follows. We call $v_1$ the lower bound on
$l_0$ and the upper bound $v_2$ ($l_0\in [v_1,v_2]$). Without
knowing which of the four entries is zero in the cell, or even
knowing the value of $m_i$, we do know from Eq. (\ref{A8}) that
$v_1\in \{0,1-m_i,m_i\}$ and $v_2\in \{1,1-m_i,m_i\}$. This
observation will be used in the ensuing subsection.

\subsubsection{A Column}

In the following we use the constraints on $l_0$ deduced in the
preceding section (Eq. \ref{A8}) to express a column of a
distribution's table as a convex combination of two simpler
columns. Recall that all cells in column $\X$ (row $\Y$) have the
same marginal $P_{a|\X}$ ($P_{b|\Y}$) by Eqs. (\ref{3},\ref{4}).
Each probability distribution has $d_\Y$ cells in the column
$\X=0$. If there is one zero in every cell of the column, there
will be $d_\Y$ overlapping ranges (see Eq. (\ref{A8})) in which
$l_0$ can lie (while keeping all other marginals, $m_i$,
constant). It is possible that $l_0$ will be uniquely determined
by these ranges (e.g. if cell $(0,1)$ requires $l_0\in[0,m_1]$ and
cell $(1,2)$ requires $l_0\in[m_2, 1]$ and $m_1=m_2$). One knows
that there is at least one value of $l_0$ consistent with all
ranges, but generically, $l_0$ will lie in a range of the form
$l_0\in [u^{(1)}_1,u^{(1)}_2]$. Here $u^{(1)}_1$ is the largest
lower bound on $l_0$ and $u^{(1)}_2$ the smallest upper bound,
with $u^{(1)}_1\in \{0,1-m_0,1-m_1...,m_0,m_1...\}$ and
$u^{(1)}_2\in \{1,1-m_0,1-m_1...,m_0,m_1...\}$. An arbitrary
distribution, $P^{(1)}$, with marginal $P^{(1)}_{a=0|\X=0}=l_0$
will have $l_0\in [u^{(1)}_1,u^{(1)}_2]$ ($u^{(1)}_1$, $u^{(1)}_2$
as defined previously). One can check that it can always be
expressed as a convex combination of two distributions $P^{(1)}_1$
and $P^{(1)}_2$ with $P^{(1)}_{1\;a=0|\X=0}=l_0=u^{(1)}_1$ and
$P^{(1)}_{2\;a=0|\X=0}=l_0=u^{(1)}_2$ respectively.

\bigskip

{\em Example:}
\begin{equation}
\label{route66}
\begin{tabular}{|ccc|}\hline
      $m_0$ && $0$ \\
       && \\
      $l_0-m_0$ && $1-l_0$ \\\hline
      $0$ && $m_1$ \\
       && \\
      $l_0$ && $1-l_0-m_1$ \\
    \hline
    $l_0$ && $m_2-l_0$ \\
       && \\
      $0$ && $1-m_2$ \\
    \hline
   \end{tabular} =\lambda\,
\begin{tabular}{|ccc|}\hline
    \;\; $m_0$  \;\; && $0$ \\
       && \\
       \;\; $0$  \;\; && $1-m_0$ \\\hline
       \;\; $0$   \;\;&& $m_1$ \\
       && \\
       \;\; $m_0$  \;\; && $1-m_0-m_1$ \\
    \hline
     \;\; $m_0$   \;\;&& $m_2-m_0$ \\
       && \\
       \;\; $0$   \;\;&& $1-m_2$ \\
    \hline
   \end{tabular}
  +(1-\lambda)\,
    \begin{tabular}{|ccc|}\hline
      $m_0$ && $0$ \\
       && \\
      $m_2-m_0$ && $1-m_2$ \\\hline
      $0$ && $m_1$ \\
       && \\
      $m_2$ && $1-m_2-m_1$ \\
    \hline
    $m_2$ && $0$ \\
       && \\
      $0$ && $1-m_2$ \\
    \hline
   \end{tabular}
\end{equation}

\bigskip

Above is a simple example of the procedure described. Without
knowing the specific values of $m_0,m_1$ and $m_2$, and without
even looking where the zeros are in each cell of the column, we do
have the basic knowledge that $l_0\in [u^{(1)}_1,u^{(1)}_2]$ where
$u^{(1)}_1\in \{0,1-m_0,1-m_1,1-m_2,m_0,m_1,m_2\}$ and
$u^{(1)}_2\in \{1,1-m_0,1-m_1,1-m_2,m_0,m_1,m_2\}$. By looking at
this specific case we can now refine our bounds on $l_0$. In the
following we suppose, as an example, that $1-m_1>m_2$. From the
cell $(0,0)$ on the left hand side of Eq. (\ref{route66}) we know,
by positivity, that $l_0\in [m_0,1]$. From the cell $(0,1)$ we
know that $l_0\in [0,1-m_1]$. From the cell $(0,2)$ we know that
$l_0\in[0,m_2]$. Taking the largest lower bound and the smallest
upper bound from these ranges, and recalling that $1-m_1>m_2$, one
finds that $l_0\in [m_0,m_2]$. The left hand side of Eq.
(\ref{route66}) can be expressed as a convex combination of two
columns where $l_0=m_0$ and $l_0=m_2$ and the $m_i$ are kept
constant. Note that each of the two columns on the right hand side
of Eq. (\ref{route66}) contains a cell which has two zeros. These
two columns each have one more zero than the column on the left
hand side.

\subsubsection{Generalizing}

In this subsection we provide a procedure which shows that any
distribution with one zero in every cell can be expressed as
convex combinations of probability distributions which are locally
equivalent to \tab II.  We first provide the loop of the procedure
and second the condition for its termination. This approach is
effectively a generalization of the decomposition of the column
given in the preceding section. From Part 1 it is sufficient to
consider only distributions which have one zero in every cell.

\bigskip

{\em Loop:} The loop considered is of the form described in Eqs.
(\ref{A1}-\ref{A5}): (I) A distribution is expressed as a convex
combination of two simpler distributions (II) one of these
distributions is selected and then becomes the distribution in
step (I) - the loop then continues.

In what follows we will follow the loop through two cycles.

\begin{description}
    \item (I) A starting distribution $P^{(1)}$ will have $l_0$ constrained to lie in a range $l_0\in
[u^{(1)}_1,u^{(1)}_2]$ where $u^{(1)}_1\in
\{0,1-m_0,1-m_1...,m_0,m_1...\}$ and $u^{(1)}_2\in
\{1,1-m_0,1-m_1...,m_0,m_1...\}$. It can be expressed
    as a convex combination of two distributions, $P^{(1)}_1$ and
    $P^{(1)}_2$.
    These satisfy the further constraints that
$P^{(1)}_{1\;a=0|\X=0}=l_0=u^{(1)}_1$ and
$P^{(1)}_{2\;a=0|\X=0}=l_0=u^{(1)}_2$ respectively. (This implies
that $P^{(1)}_1$ and    $P^{(1)}_2$ each have one cell which has
two zeros in their $\X=0$ columns.)

    \item (II) The distribution $P^{(1)}_{\ka}$, $\ka\in\{1,2\}$ is chosen as $P^{(2)}$.
    \item (I) $P^{(2)}$ will have $l_0=u^{(1)}_1$ or $u^{(1)}_2$. There is now one less parameter in the table because two of the marginals have been related. $l_0=u^{(1)}_{\ka}$ will also be constrained to lie in a new range $l_0\in
[u^{(2)}_1,u^{(2)}_2]$ where $u^{(2)}_1\in \{0,1-m_0,
1-m_1...,m_0,m_1...1-l_0, 1-l_1...,l_0,l_1...\}$ and $u^{(2)}_2\in
\{1,1-m_0, 1-m_1...,m_0,m_1...1-l_0, 1-l_1...,l_0,l_1...\}$.
$P^{(2)}$ can be written as a convex combination of a distribution
$P^{(2)}_1$, with $l_0=u^{(1)}_{\ka}=u^{(2)}_1$, and $P^{(2)}_2$
with $l_0=u^{(1)}_{\ka}=u^{(2)}_2$.
    \item (II)  The distribution $P^{(2)}_{\kb}$, $\kb\in\{1,2\}$ is chosen as $P^{(3)}$.
    \item (I) ...
\end{description}

Depending on the choices made at each step (II) the procedure
creates distributions satisfying a chain of equivalences between
their marginals:
\begin{equation}
l_0=u^{(1)}_{\ka}=u^{(2)}_{\kb}=u^{(3)}_{\kc}=...=u^{(F)}_{\ke},
\end{equation}
which will be specified by the string $(\ka,\kb,...\ke)$ with
$\kx\in \{1,2\}$. The nature of $u^{(F)}_{\ke}$ will be discussed
as part of the termination conditions.
 Noting which sets  the $u^{(i)}_1$ and  $u^{(i)}_2$ are chosen from,
  a chain of equivalences could, for example,  be of the form
$l_0=m_2=1-m_6=l_1=...$\,. Note that after each cycle the
distributions have one less parameter as more and more of their
marginals are related to each other. The procedure shrinks the
number of free parameters as it converges towards extreme points
(these have no free parameters).

The first equivalence in a  chain of equalities can only be
$l_0=m_n$ or $l_0=1-m_n$ for some $n$ (the $l_0=0$ or $1$ case
will be discussed as part of the termination conditions). This is
explained by noting that in the first cycle $l_0$ is constrained
by cells in the same column (see the preceding subsection). It
follows that
 $u^{(1)}_1$ lies in the set $\{0,1-m_0,1-m_1...,m_0,m_1...\}$ and
$u^{(1)}_2$ lies in  $\{1,1-m_0,1-m_1...,m_0,m_1...\}$ which only
depend on the values of the $m_i$. After the first cycle in which
$l_0=m_n$ or $l_0=1-m_n$, the cells in both column $(x=0)$ and the
row $(y=n)$ will provide constraints on $l_0$. This is because the
cells in row $(y=n)$ all depend on $m_n$. With some thought, one
sees that in general $u^{(j)}_1$ will thus lie in the set
$\{0,1-m_0, 1-m_1...,m_0,m_1...1-l_0, 1-l_1...,l_0,l_1...\}$ and
$u^{(j)}_2$
 in  $ \{1,1-m_0, 1-m_1...,m_0,m_1...1-l_0,
1-l_1...,l_0,l_1...\}$ and these depend on both $m_i$ and $l_i$.

\bigskip

{\em Termination conditions:} We now discuss loop termination. It
 terminates, after $F$ cycles, in two distinct ways.\begin{description}
    \item (a) When $u^{(F)}_{\ke}=0$ or $1$
    \item (b) When $u^{(F)}_{\ke}=1-u^{(g)}_{\kg}$ for $g<F$. This is only satisfied if $u^{(F)}_{\ke}=1/2$.
\end{description}

An example of case (b) would be $l_0=m_2=1-m_6=l_1=...=1-m_2$,
which implies that all of these numbers must be $1/2$.

\bigskip

 After the loop terminates, several marginals from the set of all $l_i$ and $m_i$
  will  have been set to either $0,1$, or $1/2$ (the procedure as a whole always terminates, as there are a finite number of marginals to be equated). If there exists a set of marginals which have not been fixed to one of these values,  a new marginal
$l_i$ (or $m_i$) from this set can be chosen. The form of the
above loop can then be repeated by studying constraints on this
new variable.

\bigskip

 By repeating this procedure, all marginals, $m_i$ and $l_i$, will be absorbed
into a chain of equalities terminating in $0,1$ or $1/2$. A
probability distribution equivalent to \tab II will be the only
possible result. It will generally be necessary to perform some
local relabelling to obtain distributions of the form of \tab II.
For example, the outcomes for all deterministic input settings
have to be fixed to `$0$'.


\subsection*{Part 3}
The following proves by contradiction that all distributions of
the form defined in \tab \ref{r} are extremal. Suppose that a
particular distribution $P^{(F)}_1$ of form defined in \tab
\ref{r} is not extremal. It can thus be expressed as a convex
combination of more than one distribution. Positivity requires
that these distributions have a zero where $P^{(F)}_1$ has a zero.

Suppose, from \tab \ref{r}, that $P^{(F)}_1$ has $g_\X=g_\Y=0$
then all of its cells have three zeros. This distribution cannot
be expressed as a convex combination of two distinct distributions
with the same zeros, since normalization fixes the fourth entry of
each cell to be one. $P^{(F)}_1$ is the only distribution with
these zeros.

If $P^{(F)}_1$ has $g_\X,g_\Y\geq 2$ it will have some cells with
three zeros (if $g_\X <  d_\X$  and $g_\Y < d_\Y$) and some with
two zeros. As noted above, the three zero cells have their fourth
entry fixed by normalization. A study of the distribution of zeros
in the four cells $(i,j)$, $i,j\in \{0,1\}$ shows that all
remaining non-zero entries are forced to be one-half. $P^{(F)}_1$
is the only distribution with its particular distribution of
zeros.


\begin{thebibliography}{31}
\expandafter\ifx\csname
natexlab\endcsname\relax\def\natexlab#1{#1}\fi
\expandafter\ifx\csname bibnamefont\endcsname\relax
  \def\bibnamefont#1{#1}\fi
\expandafter\ifx\csname bibfnamefont\endcsname\relax
  \def\bibfnamefont#1{#1}\fi
\expandafter\ifx\csname citenamefont\endcsname\relax
  \def\citenamefont#1{#1}\fi
\expandafter\ifx\csname url\endcsname\relax
  \def\url#1{\texttt{#1}}\fi
\expandafter\ifx\csname
urlprefix\endcsname\relax\def\urlprefix{URL }\fi
\providecommand{\bibinfo}[2]{#2}
\providecommand{\eprint}[2][]{\url{#2}}

\bibitem[{\citenamefont{{J. S. Bell}}(1964)}]{Bell}
\bibinfo{author}{\bibnamefont{{J. S. Bell}}}, \bibinfo{journal}{Physics (Long Island City, N.Y.)}
  \textbf{\bibinfo{volume}{1}}, \bibinfo{pages}{195} (\bibinfo{year}{1964}).

 \bibitem[{\citenamefont{{D.G. Collins}}(2002)}]{collinsthesis}
\bibinfo{author}{\bibnamefont{{D.G. Collins}}}, Ph.D. thesis,
  \bibinfo{school}{University of Bristol, Department of Physics}
  (\bibinfo{year}{2002}), \bibinfo{note}{available at
  http://rogers.phy.bris.ac.uk/theses.html}.

\bibitem{barrett}J. Barrett, N. Linden, S. Massar, S.
Pironio, S. Popescu and D. Roberts,  Phys. Rev. A {\bf 71}, 022101
(2005).
\bibitem[{\citenamefont{{S. Popescu} and {D.
  Rohrlich}}(1994)}]{popescu+rohrlich}
\bibinfo{author}{\bibnamefont{{S. Popescu}}} \bibnamefont{and}
  \bibinfo{author}{\bibnamefont{{D. Rohrlich}}}, \bibinfo{journal}{Found.
  Phys.} \textbf{\bibinfo{volume}{24}}, \bibinfo{pages}{379}
  (\bibinfo{year}{1994}).


  \bibitem{vandam2} W. van Dam,  e-print
quant-ph/0501159.
\bibitem{Wolf}  S. Wolf and J. Wullschleger, e-print
quant-ph/0502030.

\bibitem{Tony}  T. Short, N. Gisin and S. Popescu,
 e-print quant-ph/0504134.
\bibitem{Buhrman}H. Buhrman, M. Christandl, F.Unger, S. Wehner and
A. Winter, e-print quant-ph/0504133.

\bibitem[{\citenamefont{{J. Barrett} et~al.}()\citenamefont{{J. Barrett}, {L.
  Hardy}, and {A. Kent}}}]{qkd}
\bibinfo{author}{\bibnamefont{{J. Barrett}}}, \bibinfo{author}{\bibnamefont{{L.
  Hardy}}}, \bibnamefont{and} \bibinfo{author}{\bibnamefont{{A. Kent}}},
  \bibinfo{howpublished}{e-print quant-ph/0405101}.

\bibitem{acin} Ll. Masanes, A. Ac\'{\i}n and N. Gisin; {\em General properties of
no-signaling theories}, in preparation.
\bibitem{Cerf}  N. J. Cerf, N. Gisin, S. Massar and S. Popescu,  Phys. Rev. Lett. {\bf 94}, 220403
(2005).

\bibitem[{\citenamefont{{A. Peres}}(1990)}]{Peres}
\bibinfo{author}{\bibnamefont{{A. Peres}}}, \bibinfo{journal}{Found.
  Phys.} \textbf{\bibinfo{volume}{29}}, \bibinfo{pages}{589}
  (\bibinfo{year}{1999}).
\bibitem{cir} B.S. Tsirelson, Hadronic J. Suppl. {\bf 8}, 329
(1993).
\bibitem{Pironio} J. Barrett and S. Pironio, e-print
quant-ph/0506180.


\end{thebibliography}
\end{document}